\documentclass{mn2e}
\usepackage[cp1251]{inputenc}
\usepackage[russian]{babel}
\usepackage{graphicx,epsfig}
\usepackage{floatflt}
\usepackage {amssymb}
\usepackage {amsmath}
\usepackage {longtable}
\usepackage[usenames]{color}
\title[$N$-body simulation of clumpy torus]
    {$N$-body simulation of a clumpy torus: application to active galactic nuclei}
\author[Bannikova, Vakulik, $\&$ Sergeev]
  {E.Yu.~Bannikova$^{1,2}$\thanks {E-mail: bannikova@astron.kharkov.ua}, V.G.~Vakulik$^{1,2}$\thanks
  {Deceased} and
  A.V.~Sergeev$^{1,2}$ \\
   $1$   Institute of Radio Astronomy of the National Academy of Sciences of Ukraine, Krasnoznamennaya 4,
   61022 Kharkov, Ukraine \\
   $^2$  Karazin Kharkov National University,
   Svobody Sq. 4, 61022 Kharkov, Ukraine }

\date{Accepted 2012 April 25 .  Received 2012 April 12 ; in
original form 2012 January 18}

\pagerange{\pageref{firstpage}--\pageref{lastpage}} \pubyear{2012}

\def\LaTeX{L\kern-.36em\raise.3ex\hbox{a}\kern-.15em
    T\kern-.1667em\lower.7ex\hbox{E}\kern-.125emX}

\begin{document}

\label{firstpage}

\maketitle

\begin{abstract}
The gravitational properties of a torus are investigated. It is
shown that a torus can be formed from test particles orbiting in the
gravitational field of a central mass. In this case, a toroidal
distribution is achieved because of the significant spread of
inclinations and eccentricities of the orbits. To investigate the
self-gravity of the torus we consider the $N$-body problem for a
torus located in the gravitational field of a central mass. It is
shown that in the equilibrium state the cross-section of the torus
is oval with a Gaussian density distribution. The dependence of the
obscuring efficiency on torus inclination is found.
\end{abstract}


\section{INTRODUCTION}

Obscuring dusty tori are important structural elements of Active
Galactic Nuclei (AGN). In the framework of the unified scheme, the
differences in the AGN are explained by the orientation of the torus
relative to an observer (Antonucci 1993; Urry \& Padovani 1995).
Full obscuration of the central engine (black hole + accretion disk)
and the Broad Line Region (BLR) is realized when the torus is seen
edge-on. In this case, only  narrow emission lines are detected in
spectrum, and this type of AGN is called 'type 2'. In contrast, if
the torus is inclined at some angle to the line of sight, the
emission from the accretion disk and BLR can be seen to the observer
('type 1'). Following from the statistical data, the torus must be
geometrically thick in order to explain the observed properties of
Seyfert galaxies (Osterbrock \& Shaw 1988; Schmitt et al. 2001).

The first direct evidence for the existence of a compact structure
in the nucleus of Sy2 galaxy NGC~1068 was found with help of
near-infrared bispectrum speckle interferometry (Wittkowski et al.
1998; Weigelt et al. 2004) and speckle interferometry (Weinberger,
Neugebauer \& Matthews 1999). Subsequently, observations of
NGC~1068 using VLTI/MIDI in the infrared band ($10\mu m$) allowed,
for the first time, a spatial distribution of temperature in the
torus to be obtain (Jaffe et al. 2004). It was discovered that the
torus has two components: a hot compact component ($T > 800K$) and
a warm component ($T=300K$). The hot component is about $1.35$
parsec long and $0.45$ parsec thick; the size of the warm
component is 3 x 4pc (Raban et al. 2009). Thus, the observations
confirmed a substantial geometrical thickness of the torus: the
minor-to-major radius ratio $r_0$ is about $0.7$. It is suggested
that the hot component discovered by Jaffe et al. (2004) is the
inner funnel of the obscuring torus heated by radiation of the
accretion disc, while the warm component is the torus body itself
(Schartmann et al. 2005, Dullemond \& Bemmel 2005, Raban et al.
2009).

The mass of the obscuring torus can be estimated from the rotation
curve obtained from observations of maser emission. Greenhill et al.
(1996) obtained VLBI synthesis images of the $H_2O$ maser emission
from the central region of NGC~1068. They found a good fit to the
data by assuming a non-Keplerian rotation curve of the form
$V\propto r^{-0.31}$  on spatial scales above $0.6$ pc. The
deviation of the rotation curve from Keplerian may be a result of
the influence of the self-gravitation of the torus (Greenhill et al.
1996). It turns out that in the case of NGC~1068, the torus mass
could be comparable to the mass of the central black hole (Hur\'{e}
2002; Lodato \& Bertin 2003).

One of the problems concerning obscuring tori is how they stand up
against gravity. To explain the required geometrical thickness of a
torus, the magnitude of random speeds in the vertical direction must
be about: $\triangle V_z/V_{orb}\approx r_0$. Krolik \& Begelman
(1988) noted that if this motion is thermal, the corresponding
temperature ($\sim 10^6K$) would be far too hot for dust to survive.
Therefore it was assumed that the torus material is distributed in
clouds. To support cloud motions against losses by collisions,
orbital shear energy can be randomized if magnetic fields make the
clouds sufficiently elastic (Krolik \& Begelman 1988). However, the
required magnetic fields have not been detected. Pier \& Krolik
(1992) suggested a mechanism in which a geometrically thick torus
can be explained by the influence of the infrared radiation
pressure. Krolik (2007) obtained an explicit solution of hydrostatic
equilibrium within a torus on the assumption of a continuous medium.
In the framework of this solution it was shown that the infrared
radiation pressure can support geometrically thick structures in
AGN. In this case, the density distribution in torus cross-section
reaches its maximum near the inner edge of the torus with a
subsequent decrease to the outer edge. Schartmann et al. (2005,
2008) suggested a turbulent torus model in which the dusty clouds
are ejected by stars and take over the stellar velocity dispersion.
In this model, the torus is considered as a continuous medium and
the turbulent motions lead only to a weak stabilization against
gravity (Schartmann et al. 2010). Starburst-driven tori were
investigated with the help of a hydrodynamical simulation of the
interstellar medium by Wada \& Norman (2002) and Wada, Papadopoulos
\& Spaans (2009). The other idea is the presence of a dipole vortex
motion in the torus that can support its geometrical thickness
(Bannikova \& Kontorovich 2007). There are also models in which the
obscuring region is produced by a hydromagnetic wind (K\"{o}nigl \&
Kartje 1994; Elitzur \& Shlosman 2006). The main problem of these
models is the unknown origin of the large-scale magnetic field
needed to support such winds. An alternative to models of wind
scenarios is that of outflows driven by infrared radiation pressure
on dust (Dorodnitsyn, Bisnovatyi-Kogan \& Kallman 2011, 2012).

In hydrodynamic models the torus is considered as a continuous
medium, and hence the motion of matter (the clouds) is parallel to
the torus plane of symmetry. In this case, gravity tends to compress
the torus into a disk, and therefore the substantial forces are
required to work against gravity. In this paper we consider discrete
clouds moving in the gravitational field of the central mass. In
this case, the orbital plane of each cloud passes through the
central mass. The clouds form a Keplerian disc if their orbits lie
in one plane. The toroidal structure can be formed by clouds moving
in orbits of different inclinations and eccentricities. Such orbits
are similar to those of stars moving around the supermassive black
hole in the Galactic centre (Ghez et al. 2005). Note that recent
observations have detected in this region a dusty cloud, which also
moves in a very elongated orbit with large inclination (Gillessen et
al. 2012).

In this paper we investigate the gravitational properties of a torus
composed of clouds moving in orbits with different inclinations and
eccentricities. We start with an idealized case, considering clouds
in approximation of test particles moving on Keplerian orbits in the
gravitational field of the central mass (Section 2). In Section 3 we
considered the problem of test particle motion in the inner
gravitational potential of a homogeneous circular torus and central
mass. In Section 4, the $N$-body problem for a torus located in the
gravitational field of a central mass is investigated. The
equilibrium cross-section of self-gravitating torus is found, and
its evolution over long times is analyzed. The obtained results are
used for the interpretation of the observed features of the
obscuring tori in AGN (Section 5).

\section{KEPLERIAN TORUS}
Let us consider the problem of test particles motion in the
gravitational field of a central mass $M_c$ (Fig.1). This is a
two-body problem for each particle. We can impose conditions on the
orbital elements such that the particles form a toroidal structure.
In this case, a substantial geometrical thickness of the torus can
be reached owing to the significant scatter of inclinations and
eccentricities of the particles orbits. Because particles move on
unperturbed (Keplerian) orbits, we will call a torus formed in this
way a "Keplerian torus".
\begin{figure}
 \includegraphics[width = 84mm]{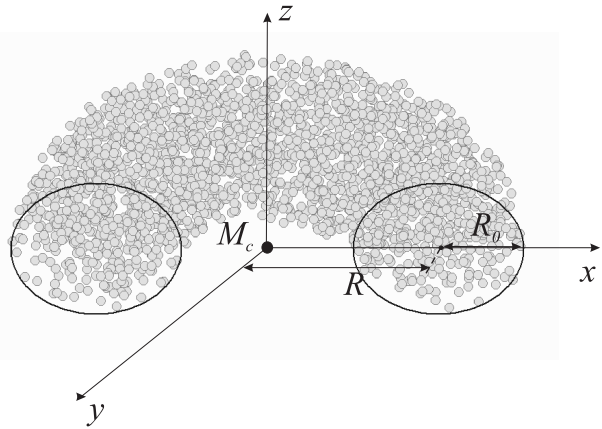}
\caption{The scheme of the torus, which consists of particles
orbiting in gravitational field of a central mass $M_c$.}
\end{figure}

To create a Keplerian torus we impose the following conditions:

a) the shape of the cross-section of the torus must be  close to
circular or elliptic;

b) the spatial distribution of particles in the torus must be  close
to homogeneous;

c) the orbits of particles in the comoving system of coordinates
must be nested. This condition allows us to avoid intersections of
orbits and hence collisions between particles.\footnote{Note that in
Section 4 we will pass from test particles to more realistic objects
that have sizes and masses. Therefore collisions between particles
will be taken into account.}

The torus is characterized by two geometrical parameters: major $R$
and minor $R_0$ radii (Fig.~1).  It is convenient to introduce the
geometrical parameter $r_0=R_0/R$.  To ensure a quasi-circular
cross-section of the torus, we should choose the semi-major axes of
the  orbits of all particles equal to the major radius of the torus.
 In order to keep all the particles under the torus surface, the eccentricities of their
 orbits must be in the range $e=[0,r_0]$ and the inclinations, measured from the symmetry
 plane of the torus, must satisfy the following condition:
\begin{equation}\label{eq2.1}
  i=\arcsin \left(
             q \frac{e}{\sqrt{1-e^2}}
             \right) ,
\end{equation}
where  $q$ is a parameter allowing us to change the ellipticity of
the torus cross-section. Let  $\rho = \sqrt{x^2 + y^2}/R$ be the
radial coordinate of a particle in the torus plane of symmetry,
measured from the central mass and normalized to $R$. Then the
coordinates of the inner and outer boundaries of the torus are
determined by the eccentricity of the orbit $\rho_{min}=1 - e$ and
$\rho_{max}=1 + e$. Let us suppose that the argument of perihelion
for all orbits $\omega = 0$; that is, the line of nodes coincides
with the semi-major axis. We generate randomly the longitude of the
ascending node $\Omega$ and the true anomaly $\nu$ in the range from
$0$ to $2\pi$. Knowing the orbit elements, we can calculate spatial
coordinates for each particle using the well-known formulae
(Duboshin 1968)
\begin{equation}\label{eq2.2}
  (x,y,z) = r \times (\alpha,\beta,\gamma),
\end{equation}
where
\[
\begin{array}{l}
\alpha = \cos\Omega\cos\nu - \sin\Omega\sin\nu\cos i \\ \beta =
\sin \Omega \cos \nu + \cos\Omega \sin\nu\cos i \\ \gamma = \sin
\nu \sin i
\end{array}
\]
and the module of the radius vector from the central mass to the
particle
\begin{equation}\label{eq2.3}
  r = \frac{R(1 - e^2)}{1 + e\cos\nu}.
\end{equation}
The velocity components of the particles are determined in the
following way:
\begin{equation}\label{eq2.4}
\left(
\begin{array}{l} V_x\\ V_y\\ V_z
\end{array}
\right) = \frac{I}{p} \left[ \left(
\begin{array}{l}
 \alpha\\ \beta\\ \gamma
 \end{array}
 \right)\cdot e \sin\nu +
\left(
\begin{array}{l}
 \alpha'\\ \beta'\\ \gamma'
 \end{array}
  \right)
(1 + e \cos \nu) \right],
\end{equation}
where
\[
(\alpha', \beta', \gamma') = \frac{d}{d\nu}(\alpha, \beta, \gamma),
\]
$p = R(1 - e^2)$ is a focal parameter, and $I = \sqrt{GM_c \cdot p}$
is a module of the kinetic moment. Thereafter, we will use the
system in which $G = R = M_c = 1$. Since $R = 1$, the value of the
geometrical parameter $r_0$ uniquely determines the minor radius of
the torus and consequently its geometrical thickness. Finally, an
algorithm to simulate Keplerian torus is the following.

1. Set the number of particles $N$, the geometrical parameter $r_0$
and the ellipticity of the torus cross-section $q$.

2. Generate randomly the eccentricities within the given limits $e =
[0, r_0]$ for each particle and find corresponding values for the
inclinations by the formula (\ref{eq2.1}). Also generate $\Omega$
and $\nu$ randomly for these orbits.

3. Find the spatial coordinates and velocity components by elements
of the orbits according to (\ref{eq2.2}) - (\ref{eq2.4}) and use
them as the initial values.

4. Solve the equations of motion with the obtained initial
conditions \footnote{The results in the form of animations are
presented at http://astrodata.univer.kharkov.ua/agn/torus/}.

Because the investigated system is symmetric with respect to
azimuthal angle, it is convenient to use a comoving system of
coordinates. This system of coordinates is formed by a plane
perpendicular to the plane of torus symmetry and passing through the
radius vector of a particle. Thus, one axis of this system $\rho$
coincides with a projection of the radius vector of a particle on
the plane of torus symmetry, while second one $\zeta = z/R$
coincides with the axis of torus symmetry. Because a particle is
orbiting around the central mass, the comoving system is also
orbiting with a velocity equal to the orbital velocity of the
particle. The comoving system of coordinates is convenient because
the trajectories of particles in this system reflect the shape of
the torus cross-section.

The equation of a particle trajectory in the comoving system of
coordinates for the case of a Keplerian torus has the following
parametric form:
\begin{equation}\label{eq2.4a}
  \rho = \frac{1 - e^2}{1 + e\cos\nu} \sqrt{1 - q^2 \frac{e^2}{1-e^2}\sin^2\nu}
\end{equation}
\begin{equation}\label{eq2.4aa}
  \zeta = q e \sqrt{1 - e^2}\frac{\sin\nu}{1 + e\cos\nu}.
\end{equation}
The trajectories of the particles in the comoving system of
coordinates calculated with (\ref{eq2.4a}) - (\ref{eq2.4aa}) are
presented in Fig~2. The shape of each orbit determines the shape of
the cross-section of the torus with a given value of $r_0$. The
period of a particle in comoving system of coordinates equals the
orbital period. Furthermore, the periods of all particles are equal
because the semi-major axes of all orbits equal unity ($R=1$).

\begin{figure}
 \includegraphics[width = 84mm]{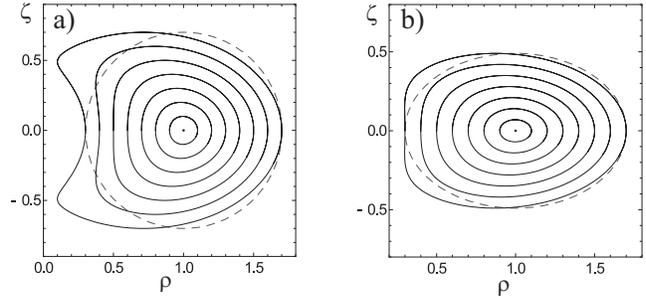}
 \caption{Trajectories of particles with the eccentricities $e=0.1 k$ ($k=1..7$)
 in a comoving system  of coordinates: a) $q=1$, b) $q=0.7$.
 The central mass is located at the point with coordinates $(0,0)$.
 Dashed lines show the circular (a) and elliptical (b) cross-sections.
 For the limiting case of a thick torus (outer solid curves),
 the cross-sections of a Keplerian torus differ noticeably from the corresponding
 circular (a) or elliptical (b) ones.}
\end{figure}

It is seen from Fig.2 that the cross-section of the Keplerian torus
is different from the circular cross-section for large values of
orbit inclinations (the outer trajectories in Fig.2). At small
inclinations, the trajectories are nearly circular  (inner curves in
Fig.2a). Interestingly, the trajectories at large inclinations make
the cross-section of the torus a more complicated shape with two
humps (Fig.2a). The restriction on the geometrical thickness of a
Keplerian torus follows from (\ref{eq2.1}). Indeed, because $\sin i
\leq 1$ we obtain the maximum value of the eccentricity $e_{max} =
1/\sqrt{q^2 + 1}$ from (\ref{eq2.1}) which is the upper restriction
on $r_0$. For $q=1$, the highest possible value of the geometrical
parameter of a Keplerian torus is $r_0 = 1/\sqrt{2} \approx 0.7$.
This limit corresponds to the maximum value of orbit inclination $i
= \pi/2$; that is, the outer particle is moving in the plane
perpendicular to the torus plane of symmetry in this case. Note that
if the eccentricities and the inclinations of the orbits of all
particles tend to  zero, then the Keplerian torus degenerates to an
infinitely thin ring. Obviously, because this torus is formed from
test particles (interaction between particles is not accounted for),
the orbits of the particles are always closed and the Keplerian
torus is stable.\footnote{Because there is an analitical solution
for $\mathbf{r}(\nu)$ in case of a Keplerian torus we use it as the
initial condition in the $N$-body simulation (Section 4).} This is a
limiting case for a self-gravitating torus when its mass tends to
zero. In reality, the torus has a mass, and, consequently, one needs
to account for its gravitational potential. In the next section we
will investigate the trajectories of a test particle in the
gravitational potential of the central mass and a homogeneous
circular torus.

\section{TRAJECTORIES OF TEST PARTICLE IN THE INNER POTENTIAL OF A HOMOGENEOUS
CIRCULAR TORUS AND THE CENTRAL MASS}

Consider the motion of a test particle in the inner gravitational
potential of both the torus and the central mass. Because the
expression for the torus potential is not integrable explicitely,
we can take advantage of its expansion for the analysis of
particle trajectories. Bannikova, Vakulik \& Shulga (2011)
obtained an approximate expression for the inner  gravitational
potential of the homogeneous circular torus in the form of a power
series up to second-order terms:
\begin{equation}\label{eq3.1}
  \varphi_{torus}^{inner} (\eta,\zeta; r_0) \approx \frac{G M_{torus}}{2\pi R}
  \left[
  c + a_1 \frac{\eta}{r_0} + a_2 \left(\frac{\eta}{r_0}\right)^2 +
    b_2 \left(\frac{\zeta}{r_0}\right)^2
 \right]
\end{equation}
where  $M_{torus}$ is the torus mass, $\eta$, $\zeta$ are
dimensionless coordinates ($\eta = \rho - 1$) and the coefficients
depend on the geometrical parameter $r_0$:
\[
\begin{array}{l}
a_1 = 8k (1+\ln k) , \qquad k \equiv r_0/8 \\ a_2 = -1 - 4k^2
(11+10 \ln k) , \qquad b_2 = -1 + 4k^2 (3 + 2\ln k).
\end{array}
\]
The approximate expression (\ref{eq3.1}) describes the inner
potential of a homogeneous circular torus with good accuracy up to
$r_0=0.5$ and allows to perform a qualitative analysis of the
trajectories in the gravitational system "torus + central mass".
The coefficient $a_1$ defines a shift of the potential maximum
with respect to the centre of the torus cross-section and the
coefficients $a_2$, $b_2$  are related to the deviation of the
torus potential from quadratic law. As noted in a previous work
(Bannikova, Vakulik \& Shulga 2011), the inner potential of the
torus can be represented as the sum of a cylinder potential and a
term comprising the curvature of the torus surface $\varphi(r) =
\varphi_{cyl} + \varphi_{curv}$. Two types of central fields are
known in which all trajectories of finite motions are closed:
$\varphi(r) \propto 1/r$ and  $\varphi \propto r^2$ (Landau \&
Lifshitz 1988). Obviously, closed orbits are possible in the
cylinder because its gravitational potential is $\varphi_{cyl}
\propto r^2$. However, the presence of a potential of curvature
$\varphi_{curv}$ brings to forming open and more complex orbits of
the particle in the inner potential of the torus. We investigate
this problem in detail. For the qualitative analysis of particle
trajectories in the inner region of the torus, we restrict
ourselves to the case of a torus with circular cross-section and
homogeneous density distribution. Since the potential of the
considered system is axisymmetric, the projection of the angular
moment of the particle on axis $\zeta$ is a constant value
($I_\zeta = Const$). In this case, it is convenient to introduce
an effective potential (Binney \& Tremaine 1994):
\begin{equation}\label{eq3.2}
  \varphi_{eff} = \varphi^{inner}_{torus} + \varphi_c - \frac{I_\zeta^2}{2(\eta +
  1)^2} ,
\end{equation}
where the potential of the central mass $M_c$
\begin{equation}\label{eq3.3}
  \varphi_c = \frac{GM_c}{R}\frac{1}{\sqrt{(\eta + 1)^2 + \zeta^2}} .
\end{equation}
We can represent $\varphi_{eff}$ in the form of a power series up
to second-order terms:
\begin{equation}\label{eq3.4}
  \varphi_{eff}(\eta,\zeta;r_0)\approx\frac{GM_{torus}}{2\pi R}
  \left[
  \tilde{c} + \tilde{a}_1 \frac{\eta}{r_0} + \tilde{a}_2 \left(\frac{\eta}{r_0}\right)^2 +
    \tilde{b}_2 \left(\frac{\zeta}{r_0}\right)^2
  \right] ,
\end{equation}
where the coefficients
\[
\begin{array}{l}
\tilde{c} = 2 \pi \mu (2 - l^2) + c \qquad \qquad \\
\tilde{a}_1 = 2 \pi \mu r_0(2 - l^2) + a_1 \qquad  \\
\tilde{a}_2 = - \pi \mu r_0^2(3l^2 - 2) + a_2 \qquad  \\
\tilde{a}_2 = - \pi \mu r_0^2 + b_2 \qquad \qquad \qquad
\end{array}
\]
$\mu = M_c/M_{torus}$ is the ratio of the central mass to the torus
mass, and the  dimensionless parameter $l$ defines a fraction of the
angular momentum from the Keplerian value $I_\zeta^2 = l^2 G M_c/R$.
It was shown in (Bannikova, Vakulik \& Shulga 2011) that the maximum
of the  potential of the homogeneous circular torus is displaced
inside from the centre of its cross-section. As a result, the torus
must be compressed along the major radius. To prevent this
compression, an orbital motion is necessary: the gravitational force
tending to compress the torus along the major radius is compensated
by the centrifugal force. This corresponds to a shift of the maximum
of the effective potential to the centre of the torus cross-section;
that is, $\tilde{a}_1 = 0$ in (\ref{eq3.4}), whence it follows a
condition on the coefficient of the angular momentum:
\begin{equation}\label{eq3.5}
  l^2 = 1 - a_1/(2\pi\mu r_0).
\end{equation}
The equations of a particle trajectory in the comoving system of
coordinates can be obtained by solving the equations of motion:
\begin{equation}\label{eq3.6}
  \ddot{\eta}\equiv \frac{d^2 \eta}{dt^2} = \frac{1}{R}\frac{\partial \varphi_{eff}}{\partial\eta};
  \qquad
  \ddot{\zeta}\equiv \frac{d^2 \zeta}{dt^2} = \frac{1}{R}\frac{\partial
  \varphi_{eff}}{\partial\zeta}.
\end{equation}
These equations have the form of a harmonic oscillator:
\begin{equation}\label{eq3.7}
  \ddot{\eta} + \omega_A^2 \eta = 0 , \qquad \ddot{\zeta} + \omega_B^2 \zeta = 0.
\end{equation}
Then the solution of the equation (\ref{eq3.7}) can be presented in
parametric form:
\begin{equation}\label{eq3.8}
\begin{array}{l}
  \eta(t) = \eta_0 \cos[\omega_A \cdot t + \alpha],\\
  \zeta(t) = \zeta_0 \cos[\omega_B \cdot t + \beta],
  \end{array}
\end{equation}
where $\eta_0$, $\zeta_0$ are the coordinates of the particle at
initial time, and
\begin{equation}\label{eq3.9}
  \omega_A^2 = \frac{GM_{torus}}{\pi R^3 r_0^2}
  \left[
  \pi \mu r_0^2(3l^2 - 2) - a_2
  \right],
\end{equation}
\begin{equation}\label{eq3.10}
  \omega_B^2 = \frac{GM_{torus}}{\pi R^3 r_0^2}
  \left[
  \pi \mu r_0^2 - b_2
  \right].
\end{equation}
Thus, the coordinates of the particle (\ref{eq3.8}) evolve like the
displacements of two harmonic oscillators with the epicycle
$\omega_A$ and vertical $\omega_B$ frequencies, respectively. In
this case the orbit is open and the trajectory in the comoving
system of coordinates fills some rectangular region during a period
$P_{box}$ whereupon this process is repeated.  These orbits are
known as box orbits (Binney \&Tremaine 1994). It is seen from
(\ref{eq3.5}) that for $\mu r_0 \gg 1$ the coefficient of angular
momentum $l \rightarrow 1$ and the ratio of epicycle frequency
$\omega_A$ to vertical frequency $\omega_B$ takes the form:
\begin{equation}\label{eq3.11}
  f \equiv \frac{\omega_A}{\omega_B} = \sqrt{\frac{\pi\mu r_0^2 - a_2}{\pi\mu r_0^2 - b_2}} .
\end{equation}
Then, the period of a box orbit, expressed in orbital periods, is
determined by the expression $P_{box} = 1/|1 - f|$. For $r_0 = 0.3$
and $\mu = 50; 25; 1/0.06$  we obtain  $P_{box} =213; 113; 80$. In
the case of  $\mu r_0 \gg 1$ ($l\rightarrow 1$)  it is apparent from
(\ref{eq3.11}) that $\omega_A \rightarrow \omega_B$. This limiting
case corresponds to the case of a Keplerian torus.

\begin{figure}
 \includegraphics[width = 84mm]{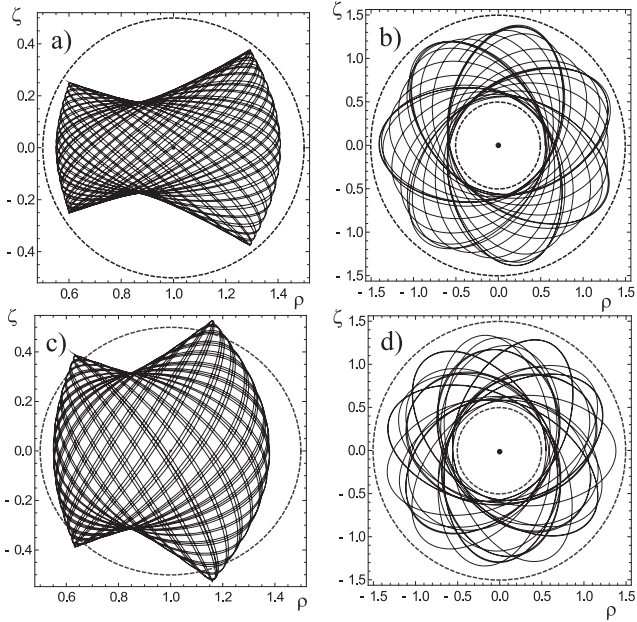}
 \caption{Trajectories of a test particle in the inner potential of homogeneous
 circular torus and the potential of the central mass. Left column: the trajectory
 in the comoving system of coordinate for  a) $V_\zeta = 0.4$; $l = 0.95$,
 c) $V_\zeta = 0.7$; $l = 0.90$.
  The centre of the torus cross-section is marked with a point. Right column:
  projection of the trajectory on the torus plane of symmetry for
  b) $V_\zeta = 0.4$; $l = 0.95$; d) $V_\zeta = 0.7$; $l = 0.90$. Dashed lines
  correspond to edges of the torus cross-section. For all cases $\mu =1$, $r_0 = 0.5$,
  $G=R=1$, $t = 150$. }
 \end{figure}

Fig.3 presents the trajectories of a test particle in the inner
gravitational potential of the homogeneous circular torus and the
potential of the central mass. For this we used an expansion for the
inner potential of the torus in a power series up to fourth-order
terms:
\begin{equation}\label{eq3.12}
  \varphi_{torus}^{inner}(\eta,\zeta;r_0) \approx \frac{G M_{torus}}{2\pi R}
  \left[
  c + \sum_{j=1} a_j \left(\frac{\eta}{r_0}\right)^j \right. +
\end{equation}
\[
  \left. + \sum_{k=1}
 b_k
  \left(\frac{\zeta}{r_0}\right)^k + \sum_{j=1}\sum_{k=1} t_{jk} \left(\frac{\eta}{r_0}\right)^j\left(\frac{\zeta}{r_0}\right)^k
  \right] ,
\]
 where the coefficients of the expansion can be found in (Bannikova,
Vakulik \& Shulga 2011). In contrast to a square-law potential
(\ref{eq3.1}), in this case we consider  a more explicit
approximation of the inner potential of the torus (\ref{eq3.12}).
Therefore,  the form of the region bounding the place of the
particle in the torus cross-section has more complex shape than
(\ref{eq3.8}) (Fig.3a,c). Note that box orbits arise as a result of
mistiming of the orbital period of the particle and its period in
the comoving system of coordinates. If the torus mass is increased,
the orbital period of particles is decreased due to turning the line
of nodes. This  leads to a shift of the pericentre and, therefore,
to the formation of a rosette orbit in the torus plane of symmetry
(Fig.3b,d). When the torus mass is decreased, the trajectory in the
comoving system of the coordinates fills the box area for a longer
time; that is, the period of the box orbit increased in this case.

\section{SELF-GRAVITATING TORUS IN THE GRAVITATIONAL FIELD OF A CENTRAL MASS: $N$-BODY SIMULATION}

This section is devoted to an investigation of a self-gravitating
torus located in the field of a central mass. We restrict
ourselves to the case of a torus whose mass is up to 10 per cent
of the central mass. As the initial condition we use a Keplerian
torus (Section 2). We investigate the following main questions:

1. Are  the tori stable on time scales comparable to the lifetimes
of astrophysical objects?

2. What is the shape of the cross-section of a torus in an
equilibrium state?

3. What is the density distribution of particles in the
equilibrium state of the torus?

As will be seen below, the self-gravitation significantly changes
the cross-section shape of the torus as compared with the case of
a Keplerian torus. The gravitational potential of  a torus with
mass $M_{torus}$, composed of $N$ gravitating particles of equal
masses, at an arbitrary point  $\bf{r}(\rho,\lambda,\zeta)$ has
the form
\begin{equation}\label{eq4.1}
  \varphi_{toru}(\rho,\lambda,\zeta) = \frac{GM_{torus}}{R}\sum_{i=1}^N \frac{1}{|\bf{r} -
  \bf{r}_i|},
\end{equation}
where $\bf{r}(\rho,\lambda,\zeta)$ is the dimensionless radius
vector of $i$-th particle normalized to the major radius of the
torus $R$. Taking into account the gravitational interaction of
particles forming the torus, the forces acting on a particle can
be represented as the sum of regular and irregular components. The
regular component is associated with a 'smoothed'  potential, and
irregular one appears because of the direct gravitational
interaction between nearest particles. The role of the irregular
forces increases in the case of a relatively small number of the
particles. Therefore, the torus potential can be represented as
follows:
\begin{equation}\label{eq4.2}
  \varphi_{torus}(\rho,\lambda,\zeta) = \varphi_{torus}^{reg}(\rho,\zeta) +
  \varphi_{torus}^{irr}(\rho,\lambda,\zeta),
\end{equation}
where the first term is the regular part of the torus potential
that is independent on the azimuth angle  $\lambda$, while the
second term is  an irregular (random) part of the potential
defined by
\begin{equation}\label{eq4.3}
 \varphi_{torus}^{irr}(\rho,\lambda,\zeta) =\frac{G}{R}
 \left[
    M_{torus} \sum_{i=1}^N \frac{1}{|\bf{r}-\bf{r}_i|} -
    \int_V \frac{\kappa(\bf{r}')dV'}{|\bf{r}-\bf{r}'|}
 \right].
\end{equation}
Here $\kappa(\bf{r})$ is the volume density of the torus.
Accounting for the central mass leads to the following expression
for the full potential:
\begin{equation}\label{eq4.4}
 \varphi(\rho,\lambda,\zeta) = \varphi_c(\rho,\zeta) +
  \varphi_{torus}(\rho,\lambda,\zeta),
\end{equation}
where the potential of the central mass is defined by expression
(\ref{eq3.3}). The limiting case $\varphi_{torus}^{irr}
\rightarrow 0$ (neglecting interactions between particles)
corresponds to the problem of the motion of a test particle in the
regular gravitational potential of the torus. This problem was
analyzed in detail in the approximation of a homogeneous circular
torus (Section 3), where it was shown that the trajectories of
particles in the cross-section of the torus relates to the type of
box orbits. In the following, we will show that the role of the
regular part of the potential appears in changing the  sizes of
the torus cross-section for $N$-body simulation. On average (by
averaging over all $i$-indices), $\varphi_{torus}^{irr}
\rightarrow 0$, but at any particular point
$\varphi_{torus}^{irr}$ can vary considerably from zero, taking
both positive and negative values. As a consequence, the velocity
of particles moving in the regular potential of the torus will be
subject to random perturbations owing to the forces determined by
the irregular part of the potential. As a measure of the irregular
part of the potential we choose a potential created by a particle
at a distance $\tilde{l} = \sqrt[3]{V_{torus}/N}$ equal to the
average distance between particles: $\varphi_i = G m_i/\tilde{l}$.
The masses of all particles are the same ($m_i = M_{torus}/N$),
and therefore $\varphi_i \propto N^{-2/3}$; that is, the irregular
forces become significant for a relatively small number of
particles $N$.

The considered $N$-body problem is reduced to the numerical
integration of the equations of motion taking account of the
central mass:
\begin{equation}\label{eq4.5}
  {\bf a}_i = -\frac{GM_c}{R^2}\frac{\bf{r}_i}{r_i^3} + \frac{\bf{F}_i}{m_i}
\end{equation}
where $\bf{a}_i$ is the acceleration of $i$-th particle. The total
gravitational force acting on $i$-th particle is
\begin{equation}\label{eq4.6}
{\bf F}_i =-\frac{G m_i}{R^2}\sum_{j=1}^N m_j \frac{\bf{r}_i -
\bf{r}_j}{\left(|\bf{r}_i - \bf{r}_j|^2 + \varepsilon^2
\right)^{3/2}},
\end{equation}
where $\varepsilon$ is a parameter allowing us to avoid the
infinite increase of the force of interaction between two
particles when they come close together (Aarseth 1963, 2003).
However, the introduction of this parameter means that we simulate
not point particles but the mega-particles of spherical form with
dimensionless radius $\varepsilon$ in which the density
distribution obeys Plummer's law (Plummer 1911; Saslaw 1985). In
the following numerical simulations we take the parameter
$\varepsilon = 0.01$. Note that the clouds penetrate through each
other for collisions in this case.

A considerable amount of computer time is needed for the numerical
simulation of a system consisting of large numbers of particles
and accounting for their gravitational interaction. In this
research we used the technology of parallel computing {\it CUDA}
developed by Nvidia, which allows us to significantly reduce the
computation time by using the computing resources of graphics
processing units (GPUs). This technology is especially effective
for $N$-body problems (Belleman, B\'{e}dorf \& Portegies Zwart
2008). The numerical simulation was performed using the graphics
card NVIDIA GeForce GTS 250 (192 {\it CUDA} Cores). The
calculation time was about 2 hours for a system of 8 192 particles
orbiting over 1000 periods with time step equal 1/150 of the
period.

Our main idea is that a torus should be more stable if it is
formed by clouds moving in inclined orbits around the central
mass. Therefore we use the Keplerian torus of the given geometric
parameter $r_0$ (Section 2) as the initial condition. The solution
for a Keplerian torus allows us to calculate the initial
coordinates and velocities of particles with inclined orbits. The
absence of quasi-periodic fluctuations of macro-parameters of a
statistical system is a necessary condition for a system of
particles to achieve a quasi-steady state. In order to test this
condition, we control the values of the kinetic energy, potential
energy and total energy of the system a few times over the period.
In addition, in order to control the dynamics of the system, the
statistical macro-parameters of the system of particles were
recorded with the same frequency, namely the coordinates of the
centre of the torus mass in its cross-section $(\eta_c, \zeta_c)$
as well as the effective sizes of the torus cross-section, which
were calculated by the following expressions
\begin{equation}\label{eq4.6}
 r_\eta = 2 \sqrt{\frac{1}{N}\sum_{i=1}^N(\eta_i - \eta_c)^2},
 \quad
  r_\zeta = 2 \sqrt{\frac{1}{N}\sum_{i=1}^N(\zeta_i - \zeta_c)^2},
\end{equation}
where $r_\eta$, $r_\zeta$ are respectively the horizontal and the
vertical sizes and the 'average' effective size of the
cross-section:
\begin{equation}\label{eq4.7}
r_{a} = \sqrt{(r_\eta^2 + r_\zeta^2)/2}.
\end{equation}
Note that in the case of a homogeneous circular disc all the
effective sizes, defined in such a way, are equal to the geometrical
radius of the disc. The angular momentum of the system of particles
in the comoving system of coordinates is defined by the expression
\begin{equation}\label{eq4.8}
  I_\theta = \sum_{i=1}^N (\eta_i \dot{\zeta}_i - \zeta_i \dot{\eta}_i),
\end{equation}
where the dot denotes a time derivative. Hereafter, this angular
momentum (\ref{eq4.8}) is referred as the 'fictitious' angular
momentum. An analysis of the behavior of the fictitious momentum
allows us to determine the degree of order (or chaos) of
directions of particles movement in the torus cross-section.
Furthermore, we periodically register the position (coordinates)
of all particles in the plane of the torus cross-section. The
accumulation of these positions during each 10 orbital periods
allowed us to obtain the average density distribution of particles
in the torus cross-section and to analyze the change of
cross-section shape during the simulation. The evolution of the
torus cross-section during the simulation for different $r_0$ is
presented in the form of an animation at
http://astrodata.univer.kharkov.ua/agn/torus.

The simulation shows that large changes of kinetic and potential
energy take place at the initial stage, $t < 20$ (Fig.4), with the
total energy conserved. The change of potential energy is related
to the change of shape of the torus cross-section  and to the
density of particles in it. Just in this short initial time
interval, the distribution of particles in the torus varies
significantly, and the torus acquires an equilibrium cross-section
of oval shape. Furthermore, the sharper part of the oval
cross-section is directed to the central mass, and the more
flattened part to the outer region. This is the opposite to the
situation in the initial state (Keplerian torus). Obviously, the
significant fluctuations in energy occur because of the tendency
of the torus to acquire a cross-section shape inverted with
respect to the initial one. For $t > 20$, the fluctuations of the
kinetic and potential energy are negligible. This means that the
shape of the torus cross-section does not change significantly
anymore. We define this state in which the shape of the torus
cross-section is no longer changing as the equilibrium state.
\begin{figure}
 \includegraphics[width = 84mm]{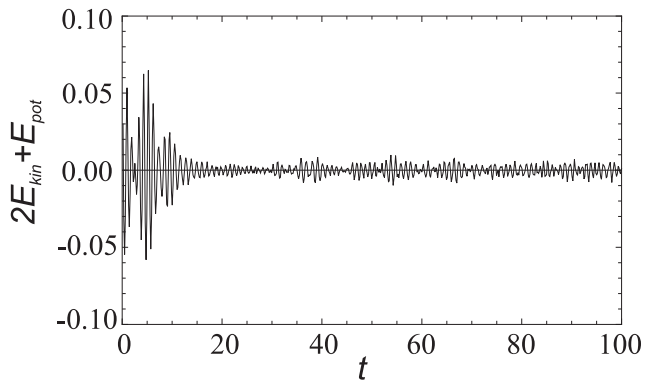}
 \caption{The sum of twice the kinetic and the potential energy of system,
 $2E_{kin} + E_{pot}$, during the first 100 orbital periods for $M_{torus}/M_c=0.0 2$ and
 $N =8~192$. Initial conditions correspond to a Keplerian torus with $r_0=0.3$.}
 \end{figure}
Fig. 4 shows that the virial theorem is satisfied, namely
$2E_{kin} + E_{pot} \approx 0$,
 which characterizes the stationary or linearly non-stationary state of the system.
\begin{figure}
 \includegraphics[width = 84mm]{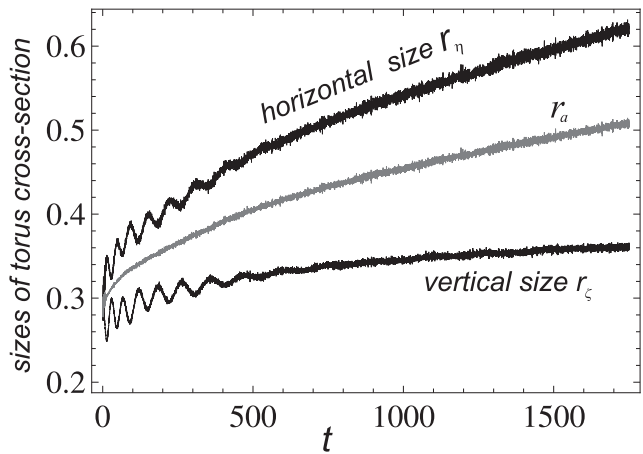}
 \caption{Variation of cross-section sizes of a torus with mass $M_{torus} =0.02M_c$,
 consisting of $N = 8~192$ particles, during 1~800 orbital periods.
 The initial conditions correspond to a Keplerian torus with $r_0=0.3$.}
 \end{figure}
Fig.5 shows that the torus cross-section gradually increases with
time. The gradual spread of the cross-section is related to the
irregular forces arising from the gravitational interaction between
particles. Indeed, the irregular forces dominate over the regular
ones caused by the smoothed potential for the considered case of a
'light' torus ($M_{torus}/M_c = 0.02$). The periodic fluctuations in
size (Fig.5) are related to the regular potential of the torus. It
was shown in Section 3 that the orbit of a test particle in the
inner potential of a circular torus in the comoving coordinate
system is of the box-orbit type. In the case of $N$-particles the
box orbits are synchronized at the initial stage, which causes
periodic fluctuations of the cross-section of the torus (Fig.5). The
amplitude of these oscillations decreases at later stages because of
the mistiming of the box orbits. As a consequence, the fictitious
angular momentum in the comoving coordinate system, $I_\theta$, also
decreases (Fig.6). The system tends to a state with $I_\theta
\rightarrow 0$, that is to the randomization  the movement
directions of particles in the torus cross-section. Note that the
total angular moment of the system is conserved.
\begin{figure}
 \includegraphics[width = 84mm]{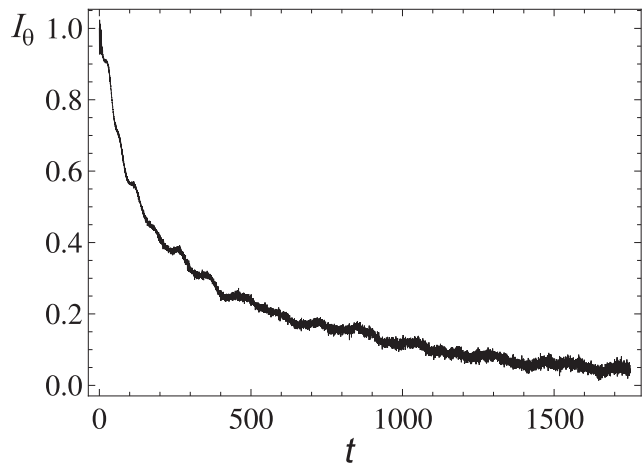}
 \caption{Variation of the fictitious angular momentum for
 a torus with mass $M_{torus} =0.02M_c$ and $N =8~192$ during 1~800
 orbital periods. $I_\theta$ is normalized to the initial value that
 corresponds to a Keplerian torus with $r_0=0.3$.}
\end{figure}
The changes of sizes of the torus cross-section for a larger value
of the initial geometrical parameter, $r_0 = 0.7$, are presented in
Fig.7. The significant difference from the previous case of a torus
with the smaller initial value $r_0 = 0.3$ is obvious (Fig.5).
\begin{figure}
\includegraphics[width = 84mm]{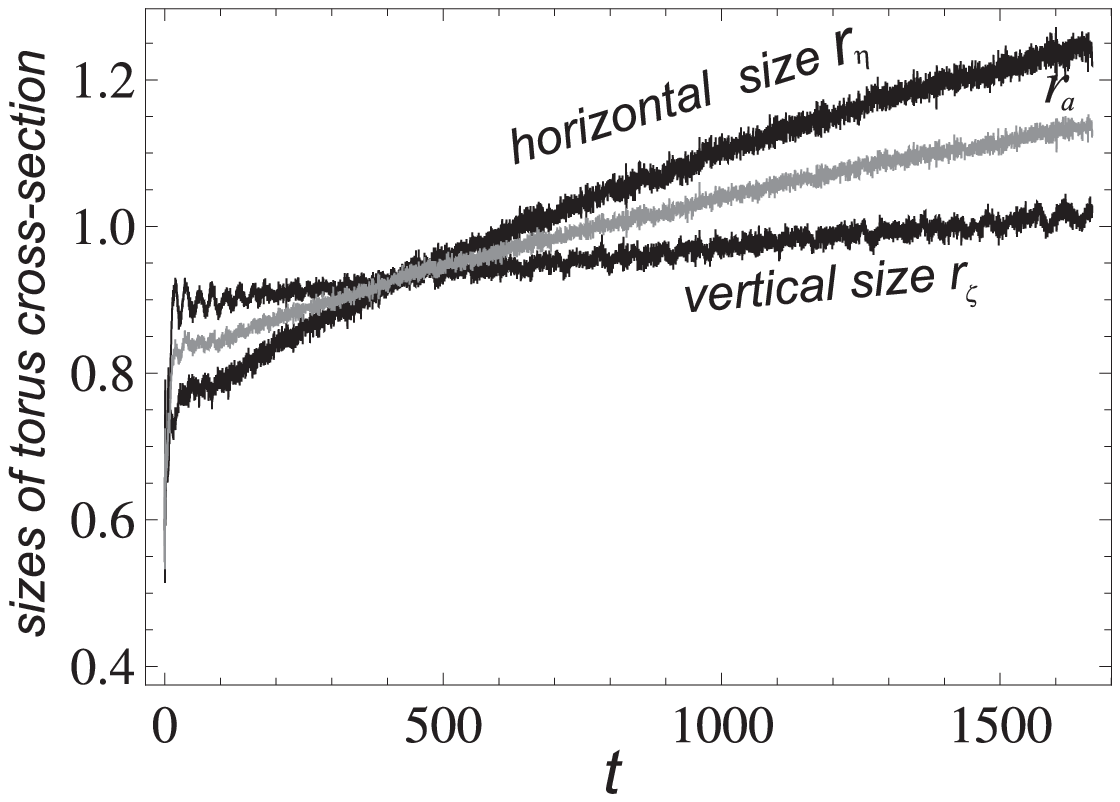}
 \caption{Variation of cross-section sizes of a torus with mass
 $M_{torus} =0.109 M_c$, consisting of $N = 8~192$ particles,
 during 1~800 orbital periods. The initial conditions correspond to
 a Keplerian torus with $r_0=0.7$.}
 \end{figure}
 Namely, the sizes of the cross-section change considerably during
 the initial time, which is connected to a significant change in the shape
 of the torus cross-section. Subsequently, the amplitude of the
 fluctuations decreases and finally disappears after $t = 300$.
 This behavior can be explained as follows. In the case of a thick torus,
 the periods of box orbits in the initial stage are very different, which
 leads to a rapid mistiming of orbits in the torus cross-section.
 This confirms the analysis of change in the fictitious angular momentum
 (Fig.8).
\begin{figure}
\includegraphics[width = 84mm]{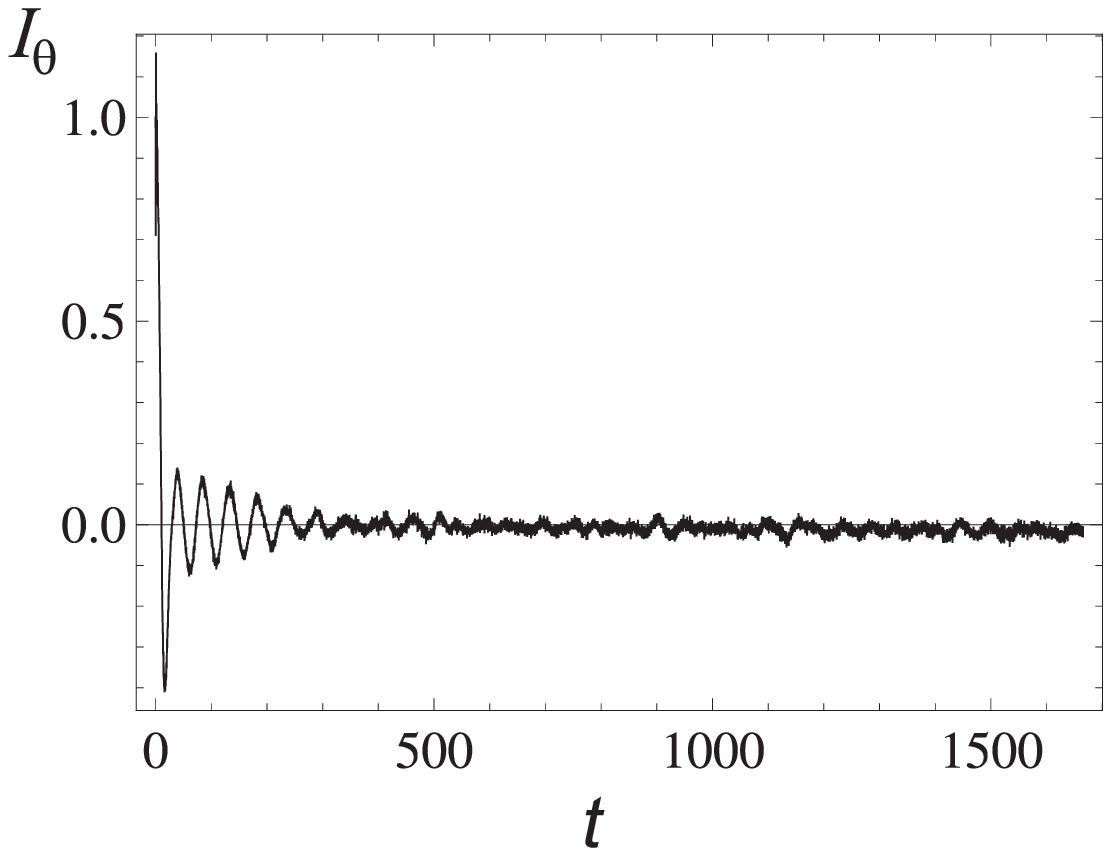}
 \caption{Variation of fictitious angular moment for a torus with
  mass $M_{torus} =0.109 M_c$ and $N = 8 192$ during 1800 orbital periods.
 $I_{\theta}$ is normalized to the initial value that corresponds to a Keplerian
 torus with $r_0=0.7$. }
 \end{figure}
 Thus, the cross-section of a self-gravitating torus spreads slowly
 (here with the vertical size increasing more slowly than the horizontal one).
 In the equilibrium state, the directions of particle movements in
 the torus cross-section are randomized.
  Note that if the central mass
 $M_c = 10^7 M\odot$  then one orbital period ($t = 1$) in the
 centre of the cross-section of a torus with the major radius $R = 2pc$ correspond
 to an average interval of $80~000$ years (see Section 5). Thus,
 the results of numerical simulation for $N=8~192$ (Figs.5-8) for
 a time interval of $1~700$ orbital periods correspond to 136 million years.

 When particles interact with each other, some of them lose energy and move
 to lower orbits, while the energy of other particles increases and they shift
 to more distant orbits or even escape the system. As a result of such
 interactions, the outer boundary of the torus cross-section gradually
 increases. By virtue of the random nature of particle interactions
we can assume that the modulus of the total change of a particle
velocity $\triangle V$ is related to velocity changes in separate
encounters of particles $\triangle V_i$ as
\begin{equation}\label{eq4.9}
  (\triangle V)^2 = \sum_i(\triangle V_i)^2 .
\end{equation}
In the theory of stellar dynamics there is well-known expression for
an estimate of the velocity dispersion  $(\triangle V)^2$ reached
during the interval $\triangle t$ (Kulikovskii 1978, Saslaw 1985):
\begin{equation}\label{eq4.10}
 (\triangle V)^2 = 32\pi G m^2 \triangle t \cdot D \left(\overline{\frac{1}{V}}\right)
 \ln \left(\frac{b_{max}}{B}\right),
\end{equation}
where $m$ is  mass of a particle, $D$ is  number of particles per
unit volume, which for a homogeneous distribution is proportional to
$N$. Omitting the constant coefficients and assuming that the mass
of particles $m = M_{torus}/N$, the expression (\ref{eq4.10}) takes
the form
\begin{equation}\label{eq4.11}
 (\triangle V)^2 \propto G \frac{M_{torus}^2}{N}\triangle t \ln\left(\frac{b_{max}}{B}\right).
\end{equation}
As an estimate of the maximum impact parameter $b_{max}$, the
average distance $\tilde{l}$ between particles is commonly used.
Then, the expression for the impact parameter is
\begin{equation}\label{eq4.12}
  b_{max} \approx \bar{l} = \sqrt[3]{\frac{2\pi^2 R^3 r_0^2}{N}}
\end{equation}
and
\begin{equation}\label{eq4.13}
  B = \frac{2G M_{torus}}{N V^2}.
\end{equation}
In the considered dimensionless system we obtain the estimate
\begin{equation}\label{eq4.14}
  \ln\left(\frac{b_{max}}{B}\right) \approx \ln\left(\frac{N^{2/3}}{M_{torus}^2}\right).
\end{equation}
Neglecting the weak logarithmic dependence $(\triangle V)^2$  of
$N$ and $M_{torus}$ we finally obtain
\begin{equation}\label{eq4.15}
  (\triangle V)^2 \sim M_{torus}^2 \frac{\triangle t}{N}.
\end{equation}
Although the expression (\ref{eq4.15}) was obtained for a simply
connected system of particles, it works satisfactorily in the case
of a torus, which is confirmed by results of simulation. For
example, if $N_1$ is number of particles in the torus and $\triangle
t_1$ is interval during which a certain value of the velocity
dispersion is reached, then it follows from (\ref{eq4.15}) that
$\triangle t_2 = \triangle t_1 \cdot N_2/N_1$; that is, the same
value of the velocity dispersion for $N_2= N_1/2$ will be reached
during $\triangle t_2= \triangle t_1/2$. Variations of cross-section
sizes for tori with different numbers of particles are shown in
Fig.9. Black curves were obtained from simulation of a torus with
$N_1=16~384$  for the interval $t_1 = 3000$. The gray curves
correspond to the torus that consist of $N_2 = N_1/2=8~192$
particles for interval $\triangle t_2= \triangle t_1/2=1~500$ with
subsequent expansion of the obtained curves by a factor of two. The
black and gray curves  coincide well, which indicates that the
expression (\ref{eq4.15}) is satisfactory.
\begin{figure}
 \includegraphics[width = 84mm]{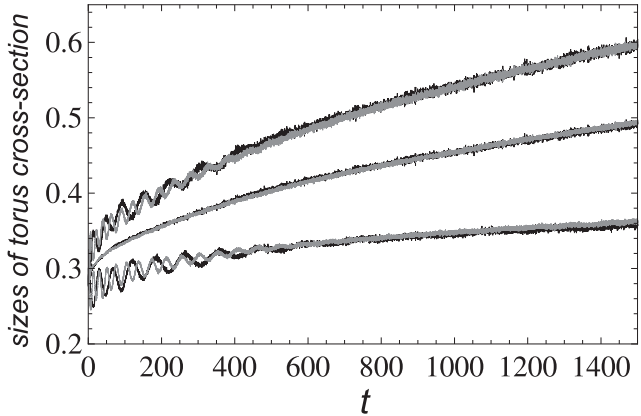}
 \caption{Variation of cross-section sizes of a torus with
 initial conditions $r_0=0.3$, $M_{torus}=0.02M_c$. The black curves
 are obtained by simulation with  $N_1=16~384$ for the interval
 $\triangle t_1 = 3000$. The gray curves correspond to case
 $N_2=N_1/2=8~192$ at $\triangle t_2 = 1~500$ with subsequent
 expansion of the obtained curves by factor of two.}
 \end{figure}
{\it Thus, provided we obtain results of the simulation of the
system with a smaller number of particles in a relatively short time
interval, expression (\ref{eq4.15}) allows us to predict the state
of the system consisting of a larger number of particles for larger
time intervals. In a similar way,  the condition of re-scaling
(\ref{eq4.15}) can be used to investigate systems with different
values of $M_{torus}$.}

We discovered from simulations that in the equilibrium state the
torus has an oval cross-section with a density distribution that
changes exponentially. The stability of the torus is achieved
because the clouds move on inclined orbits. The distribution of
particles in the torus cross-section was obtained by averaging their
coordinates on the interval of 100 orbital periods with a subsequent
approximation by a function of the form
\begin{equation}\label{eq4.16}
  n(\eta,\zeta;r_0) = n_0(r_0)\exp[-f(\eta,\zeta;r_0)],
\end{equation}
where $n_0$ is the number of particles at a maximum of the
distribution. For simplicity we set $n_0 = 1$. Obviously,
$\ln(n/n_0) = f(\eta,\zeta)$ is equation of an isodence at level
$n/n_0$ for a given density distribution (\ref{eq4.16}). Because the
isodences have an oval shape we approximate the function $f$ as a
power series in the coordinates:
\begin{equation}\label{eq4.17}
\begin{array}{l}
  f(\eta,\zeta) = c + a_1 \eta + a_2 \eta^2 + b_2 \zeta^2 + a_3 \eta^3 + \\
  + t_{12}\eta\zeta^2 + a_4 \eta^4 + b_4 \zeta^4 + t_{22}\eta^2\zeta^2 .
\end{array}
\end{equation}
Fig.10 shows an average  density distributions in the equilibrium
cross-section of a torus consisting of  $N = 8~192$ particles for
various values of the torus mass and various initial $r_0$. These
cross-sections were achieved in 1000 orbital periods. The parameters
($M_{torus}$ and $r_0$) in Fig.10 are chosen so that
 in all cases the tori have the same values of the initial volume density:
$\kappa_{torus} = M_{torus}/(2 \pi^2 R^3 r_0^2) = Const$. We also
performed special simulations by changing the initial shape of the
torus cross-section. The results of the simulations showed that
changes of the initial shape of the torus cross-section (the
ellipticity parameter $q$ in (\ref{eq2.1})) does not substantially
influence the equilibrium shape of the cross-section.
\begin{figure*}
 \includegraphics[width = 170mm]{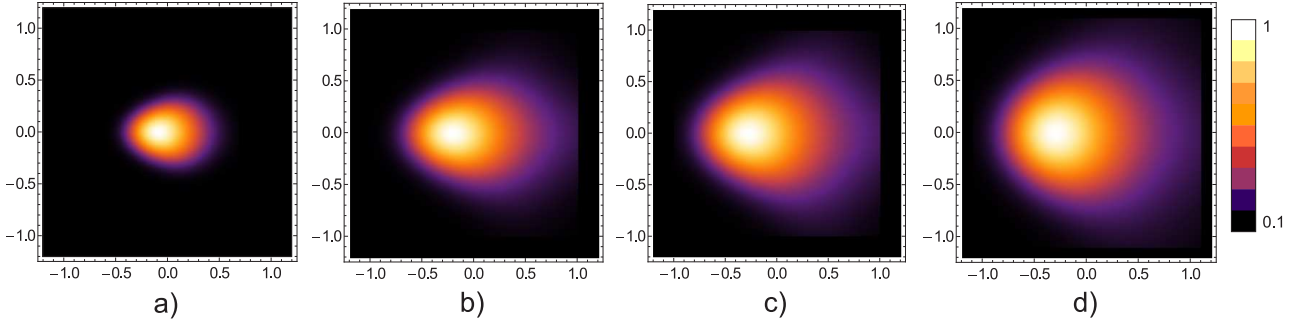}
 \caption{An average density distribution of particles in an equilibrium cross-section of torus
 consisting of  N=8 192 particles for various values of the torus mass and
 the initial geometrical parameter:
 a)$r_0 = 0.3$, $M_{torus} = 0.02M_c$, b)$r_0 = 0.5$, $M_{torus} = 0.056M_c$
 c) $r_0 = 0.6$, $M_{torus} = 0.08M_c$, d) $r_0 = 0.7$, $M_{torus} = 0.109M_c$.
 The density distributions are plotted in cylindrical coordinates ($\eta$; $\zeta$) and
 the central mass is at the point with coordinates (-1, 0).}
 \end{figure*}

So, we come to the following conclusion.

{\it 1. A gravitating torus with mass $\lesssim 0.1M_c$ is stable on
time-scales comparable to the lifetimes of the astrophysical
objects.

2.  In the general case, the equilibrium cross-section of the torus
has an oval shape. In equilibrium state, the directions of particle
movements in the torus cross-section are randomized.

3.  The density distribution of particles in the torus cross-section
obeys Gauss' law with the maximum near to the centre of the
cross-section.}

In these simulations we assumed that clouds do not change their
shape for close approaches and collisions. Real cloud could
significantly change their shape as a result of tidal shearing and
collisions. Therefore the question arises: how can orbiting dusty
clouds survive for that long time? Krolik \& Begelman (1988)
considered this problem in detail. They argued that tidal shearing
can shred the clouds into smaller pieces. On the other hand, during
collisions the material of clouds can be gathered into one giant
clump. A steady state of the clouds can exist if the rates of the
two processes are comparable (Krolik \& Begelman 1988).  Our point
is that on average these processes do not change considerably the
spread of orbits inclinations. Wind could be an additional source of
clouds moving on such orbits: the clumpy wind coming off the
accretion disk can feed the torus (Nenkova et al. 2008 Part~II). In
this case such clouds could acquire inclined orbits.

 \section{APPLICATION TO THE OBSCURING TORI IN AGN}

The obtained results (cross-section shape and density distribution)
can be directly applied to the obscuring tori of AGN. We will use
the condition of obscuration obtained from analysis of the spectral
energy distribution: for total obscuration of the central engine,
the number of clouds in the equatorial plane must be greater than 5
(Nenkova et al. 2008 Part~I,II).

\subsection{Obscuration of the central engine by the torus}

Let us find the restrictions on size and number of clouds from a
requirement of total obscuration in the equatorial plane. Assuming
an optical depth of clouds in the ultraviolet and visible light much
greater than unity, we shall consider them as an opaque slab with an
area of cross-section  $\pi\varepsilon^2R^2$, where the effective
radius of a cloud is $R_{cl} = \varepsilon R$ . The volume per one
cloud in the case of a circular torus with a homogeneous
distribution of density has the form
\begin{equation}\label{eq5.1}
  V_N = \frac{V_{torus}}{N} = \frac{2\pi^2 R^3 r_0^2}{N} .
\end{equation}
We determine the average number of clouds $N_s$ along the line of
sight in the equatorial plane of the torus as a product of the
number density $1/V_N$ and the volume of a cylinder with the
cross-sectional area $\pi\varepsilon^2 R^2$ and cylinder length
equals to the diameter of the torus cross-section
\begin{equation}\label{eq5.2}
  N_s = \frac{N}{V_{torus}}\pi\varepsilon^2R^2 2R r_0 = \frac{N\varepsilon^2}{\pi r_0} .
\end{equation}
However, the cross-section of a torus affected by self-gravitation
takes an oval shape(Fig.10) and the distribution density of the
clouds in its cross-section is inhomogeneous (Section 4). Therefore,
the expression for the average number of clouds along the line of
sight  will differ from (\ref{eq5.2}). In this case, the value of
the normalization factor $n_0$ in (\ref{eq4.16}) can be determined
from the following condition: the total number of clouds in the
torus remains constant when the cross-section of the torus is
changed owing to self-gravitation; that is,
\begin{equation}\label{eq5.4}
N = 2\pi n_0 \iint \limits_S dS \exp[-f(\eta,\zeta;r_0)].
\end{equation}
Here the integral is taken over an area of torus cross-section $S$.
To obtain the dependence of the number of clouds on the angle
$\theta$ between the equatorial plane and the line of sight we
transform into the polar system of coordinates with the origin at
the central mass: $\rho = \eta + 1 = r\cos \theta$, $\zeta = r\sin
\theta$. Then the number of clouds along the line of sight as a
function of angle is
\begin{equation}\label{eq5.4}
  N_s(\theta) = \frac{N \varepsilon^2}{2}\frac{\int\limits_0\limits^\infty e^{-f(r,\theta)}dr}
                  {\iint\limits_S e^{-f(\eta,\zeta)}dS}.
\end{equation}
Thus, the average number of clouds along the line of sight in the
equatorial plane of the torus,  taking into account the
inhomogeneous distribution of particles (\ref{eq5.4}), is given by:
\begin{equation}\label{eq5.5}
  N_s = N_s(0) = \frac{N \varepsilon^2}{2}\frac{\int\limits_0\limits^\infty e^{-f(r,0)}dr}
                  {\iint\limits_S e^{-f(\eta,\zeta)}dS}.
\end{equation}
For the limiting case of a homogeneous circular torus, the integral
in the numerator is equal to $2r_0$  and the integral in the
denominator is equal to the area of a circle, and thus the
expression (\ref{eq5.5}) reduces to (\ref{eq5.2}).
\begin{figure}
\includegraphics[width = 84mm]{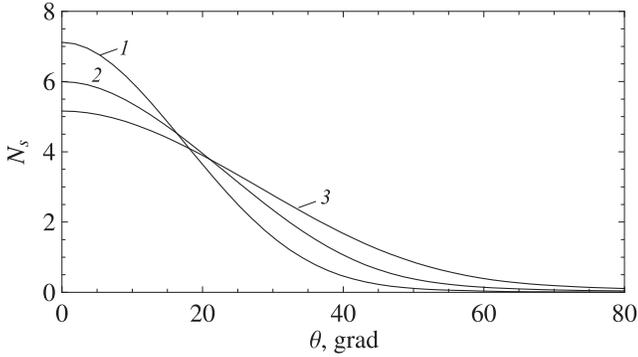}
 \caption{Dependence of the number of clouds along the line of sight on
 the equatorial angle $\theta$ calculated by (\ref{eq5.4}). The different
 curves correspond to tori with equilibrium cross-sections obtained
for different values of the torus mass and
 the initial geometrical parameter:
 1)$r_0 = 0.5$, $M_{torus} = 0.056M_c$
 2) $r_0 = 0.6$, $M_{torus} = 0.08M_c$, 3) $r_0 = 0.7$, $M_{torus} =
 0.109M_c$.}  The total number of clouds in the torus $N = 10^5$ and $\varepsilon = 0.01$.
 \end{figure}
The number of clouds along the line of sight is an important
parameter because it characterizes the optical thickness of the
torus. Indeed,  by virtue of the random distribution of clouds in
the torus, the fraction of non-obscured area $I_p$, which
corresponds to the fraction of transmitted radiation of the central
engine, is determined by a Poisson distribution $I_p = prob(k)=N_s^k
e^{-N_s}/k!$ with $k = 0$, where $k$ is the frequency of events.
Consequently, $I_p = e^{-N_s}$. On the other hand, because $I_p =
e^{-\tau_V}$, the optical depth of the torus in visible light equals
the average number of clouds: $\tau_V = N_s$. Fig.11 shows the
number of clouds along the line of sight as a function of the
equatorial angle $\theta$ for three cases corresponding the
equilibrium cross-sections in Fig.10 b,c,d. In order to fit the
observational data, the number of clouds along the line of sight in
the equatorial plane of the torus $N_s$  should be greater than 5
(Nenkova et al., part 2,  2008). If $\varepsilon = 0.01$ then this
condition holds for a torus with total number of clouds $N = 10^5$
and $r_0 \geq 0.5$. The total number of clouds decreases with the
increasing relative size of clouds because $N\propto
\varepsilon^{-2}$.

\subsection{Dynamics of clouds in the torus}

Clouds in the torus are orbiting in the gravitational field of the
central mass. Owing the fact that the orbits of clouds have a
scatter in eccentricities and inclinations, they form a toroidal
structure. Although the orbits are perturbed by mutual attraction of
the clouds, the average velocity near the torus plane of symmetry
can be determined as the velocity in a circular orbit:
\begin{equation}\label{eq5.12}
  V \simeq 210 km/s
  \left(
    \frac{M_c}{10^7 M_{\odot}}
  \right)^{1/2}
  \left(
    \frac{r}{1pc}
  \right)^{-1/2}.
\end{equation}
Then the orbital period of a cloud can be estimated as
\begin{equation}\label{eq5.19}
  P \simeq 30~000 year \cdot \left(
    \frac{M_c}{10^7 M_{\odot}}
  \right)^{-1/2}
  \left(
    \frac{r}{1pc}
  \right)^{3/2}.
\end{equation}
For the central mass $M_c =10^7 M_{\odot}$ the velocity is $V = 150
km/s$ at distance $r = 2pc$, and the corresponding orbital period $P
\simeq 80 000 year$. The inner edge of the torus ($r_{min}$) is
formed by clouds that are moving in elliptical orbits and pass
through the pericentre, while the outer edge ($r_{max}$) is formed
by clouds that pass through the apocentre. So, the cloud velocities
in these two regions can be estimated approximately as:
\begin{equation}\label{eq5.13}
  V_{\max} \simeq V_{0}\sqrt{\frac{1+r_0}{1-r_0}} ,
  \qquad V_{\min} \simeq V_{0}\sqrt{\frac{1-r_0}{1+r_0}},
\end{equation}
where $V_0=\sqrt{GM_c/a}$ and $a = (r_{max}+r_{min})/2$ is the
semi-major axis.  For example if $r_{min}=0.4 pc$ and $r_{max}=4 pc$
the maximum velocity of clouds  $V_{\max} \simeq 330 km/s$ and the
minimum velocity $V_{\min} \simeq 60 km/s$ for $r_0 = 0.7$.

\section{Conclusion}
In this paper the gravitational properties of a torus have been
investigated. It is shown that the presence of a central mass is a
necessary condition for the stability of a self-gravitating torus.
If the mass of the torus is much lower than the central mass, such a
system can be considered in the framework of the problem of test
particles motion in the central gravitational field. The toroidal
distribution of particles is achieved as a result of the significant
spread of inclinations and eccentricities of their orbits. In this
case the particles move on Keplerian orbits, and we call a torus
formed in this way as 'Keplerian torus'. It is shown that Keplerian
torus has a limit on the geometric parameter.

To investigate the properties of a self-gravitating torus we
considered the $N$-body problem for a torus located in gravitational
field of a central mass. As initial conditions we used the Keplerian
torus. It is shown that the self-gravitating torus is stable on
time-scale comparable to lifetimes of astrophysical objects. The
equilibrium cross-section of the torus has an oval shape with
Gaussian density distribution. Until now, the cross-section shape
and the distribution of particles in a self-gravitating torus were
not known and therefore were taken as free parameters in problems
such as radiation transfer in an obscuring torus in AGN.  We found
the dependence of the obscuring efficiency as a function of the
angle between the line of sight and the torus plane of symmetry.

An account of the dissipation processes is a separate issue that is
beyond the scope of this paper. The clouds can be heated as a result
of collisions, thus creating additional radiation pressure inside
the torus. It is likely that these processes can lead to a
'repulsion' of clouds and hence additionally influence at the shape
of the torus cross-section.

\section*{Acknowledgments}

This work was partly supported by the National Program
"CosmoMicroPhysics".

We are grateful to ananonymous referee for the most valuable
comments and suggestions, and to Peter Petrov for helpful
discussions.

\end{document}